\documentclass[reprint,longbibliography,pre,amsmath,amssymb,amsfont, url]{revtex4-2}

\usepackage[colorlinks, allcolors=blue]{hyperref}
\usepackage{graphicx} 
\usepackage[inline]{enumitem}
\usepackage{eurosym}
\usepackage[cal=dutchcal]{mathalfa}
\usepackage{csquotes}
\usepackage{mathtools}

\usepackage{tikz}
\usetikzlibrary{arrows, fit}

\newcommand{\myquote}[1]{\textit{\textquote{#1}}}
\newcommand*\diff{\mathop{}\!\mathrm{d}}

\begin{document}
	\author{D. Lairez}
	\email{didier.lairez@polytechnique.edu}
	\affiliation{Laboratoire des solides irradi\'es, \'Ecole polytechnique, CEA, CNRS, IPP,
		91128 Palaiseau, France}
	\title{On the supposed mass of entropy and that of information}
	\date{\today}
	
\begin{abstract}
	In the theory of special relativity, energy can be found in two forms: kinetic energy and rest mass. Potential energy of a body is actually stored under the form of rest mass, interaction energy too, temperature is not.
	Information acquired about a dynamical system can be potentially used to extract useful work from it.
	Hence the \textquote{mass-energy-information equivalence principle} that has been recently proposed.
	In this paper, it is first recalled that for a thermodynamic system made of non interacting entities at constant temperature, the internal energy is also constant. So that, the energy involved in a variation of entropy  ($T\Delta S$) differs from a change in potential energy stored or released and cannot be associated to a corresponding variation of mass of the system, even if it is expressed in term of quantity of information.
	This debate gives us the opportunity to deepen the notion of entropy seen as a quantity of information, to highlight the difference between logical irreversibility (a state dependent property) and thermodynamical irreversibility (a path dependent property) and to return to the nature of the link between energy and information that is dynamical.
 \end{abstract}
	
\maketitle


\section*{Introduction}

The link between information and energy finds its origin in Maxwell's demon who, by acquiring information about a thermodynamic system, is able to act on it and produce work in return\,\cite{Maxwell_1872}. 
Later, Shannon\,\cite{Shannon_1948} formalized mathematically this link by considering the quite different problem of information processing. He demonstrated that the minimum average number $H$ of bit to encode a random variable emitted by a source, let say the current microstate of a dynamical system, is equal to a factor $\ln 2$ to the Gibbs entropy $S$, that is itself equal to the Clausius entropy of the system: $S  =  H \ln 2$ (in this paper $S$ is dimensionless and temperature $T$ is in Joule). Hence the link: from the second law of thermodynamics, acquiring one bit of information about a dynamical system has a minimum energy cost equal to $T\ln 2$ that can in return be potentially used to extract at best the same quantity of energy from the system.

Landauer followed by Bennett\,\cite{Landauer_1961, Bennett_1982, Bennett_2003} tackled the problem in a quite different way. In their spirit, the logical states 0 or 1 of one bit of information correspond necessarily to two different thermodynamic states. Even more, any irreversible logical operation, such as erasing one bit, corresponds to an irreversible non-quasistatic thermodynamic process that consequently has a non-zero minimum energy cost when performed cyclically. This is the so called \textquote{Landauer principle}.
In this way, it is believed that \myquote{Information is physical}\,\cite{Landauer_1991} in a much convincing manner than with Shannon's information theory.

Based on the Landauer-Bennett idea, a new step (in a wrong direction) has recently been done. Information stored under the form of physical bits is considered a kind of potential energy to which, in the framework of special relativity, it can be assigned a mass\,\cite{Vopson_2019, Vopson_2020, Vopson_2022, Masic_2021}. This is the \textquote{mass-energy-information equivalence principle} that
\myquote{states that information is a form of matter, it is physical, and it can be identified by a specific mass per bit while it stores information... It is shown that the mass of a bit of information at room temperature (300K) is $3.19 \times 10^{-38}$\,Kg.}\,\cite{Vopson_2019}.
This idea has been already criticized\,\cite{Burgin_2022} at an epistemological and ontological level: what exactly does \textquote{physical} mean in \myquote{Information is physical}?\,\cite{Landauer_1991}.
The aim of this paper is to show that this idea is also false for at least three reasons, which this time are at a more prosaic level.

In the first section, the \textquote{mass-information equivalence principle}, is addressed from the thermodynamic side as a \textquote{mass-entropy equivalence principle}.
It is a recall of the basic difference between potential energy and entropy: the elastic energy of a spring is fundamentally different from that of a rubber or from that of a compressed volume of gas. For a spring it originates from a microscopic interaction potential, whereas it is emergent for a rubber or a gas.
It will be shown that a monothermal variation of entropy ($T\Delta S$) of a body is not accompanied by any variation of its mass.

In the second section, the \textquote{mass-energy-information equivalence principle} is addressed at its root, that is to say the Landauer principle.
In a previous paper\,\cite{Lairez_2023}, it has been shown that the Landauer-Bennet idea cannot be a general principle but is only true in a particular case. It follows that any derivative of this \textquote{principle} is logically ruled out. Here, new examples will be given to illustrate that logical and thermodynamical irreversibilities are uncoupled. In fact, as defined by Landauer himself, the logical irreversibility of an operation is intrinsic to its initial and final states and is independent of the path used to achieve the operation, contrary to the thermodynamic irreversibility that is a property of the path.

In the third section, the last argument against the \textquote{mass-energy-information equivalence principle} is given: the link between information and energy is valid for fresh information about a dynamical system. Old information, or information detached from its subject matter, is no longer information and has no value.

\section{Potential versus entropic forces}\label{pe_vs_e}
	
After Shannon\,\cite{Shannon_1948}, we know that the thermodynamic state quantity $S$, named entropy and introduced by Clausius\,\cite{Clausius_1879} to account for exchanges of heat during a process, is to a factor $\ln 2$ mathematically equal to the minimum average number $H$ of bits necessary to encode in which microstate the dynamical system is currently, namely the quantity of information emitted by the system. 
\begin{equation}\label{Shannon1}
	S  =  H \ln 2 
\end{equation}
Even if Shannon's information theory is not used by Landauer and Bennett, they do not question its correctness.
It follows that the hypothetical \textquote{mass-information equivalence} is nothing other than a \textquote{mass-entropy equivalence} that can be addressed in a pure thermodynamic framework in the context of special relativity. This is the aim of this section.

In the theory of special relativity, the energy of a body takes two forms: kinetic energy and mass (or rest mass, or rest energy). Mass is the energy that is stored by the body when it is at rest. For any monothermal transformation, the product of temperature $T$ (in Joule) to the variation of entropy $\Delta S$ (dimensionless) has the dimension of an energy. Behind the idea of a \textquote{mass-entropy equivalence} is that $T\Delta S$ is a sort of potential energy stored (or released) somewhere in the system at the end of the transformation, to which it can be attributed an equivalent mass difference $T\Delta S /c^2$ (where $c$ is the celerity of light), in virtue of the Einstein famous equation. Before to address this analogy, let us first deal with the case of potential energy in mechanics.

\subsection{Mechanical potential energy}

When a force is applied to a body over a given distance, mechanical work is done, that is to say energy is transferred from one body (the one applying the force) to another (the one we are interested in).
But an important point is that \myquote{Work is a process; once done it no longer exists. It is something that cannot be stored; what is stored is energy}\,\cite{Hecht_2019b}.

When a stone is transported from the ground to a table, mechanical work is done against the force of gravity. 
The energy transferred to the stone is recovered under the form of kinetic energy when the stone falls back down. If energy is conserved, where is it between these two processes? We usually say that it is stored under the form of potential energy in the earth-stone system.
But as noted by Hecht\,\cite{Hecht_2016}, kinetic energy can be measured, as well as the work that has been done, without affecting their integrity, but the potential energy of the stone on the table cannot. When we measure it, it is no longer potential energy, it is kinetic energy. 
Potential energy is a concept that was introduced to ensure the conservation of energy, the energy is actually stored under the form of mass, a physical quantity that can in principle be measured without affecting its integrity. For instance, the mass equivalence of potential energy can be measured for nuclear fission: the mass of a nucleus is smaller than the sum of those of its nucleons taken independently. The difference is due to the attractive strong interactions between nucleons and is divided between the different parts when they separate.
Even if it is not measurable for a stone on a table, for the consistency of the theory
we are obliged to assume the same effect.
The stone has more mass on the table than on the ground.

\begin{figure}[!htbp]
	\begin{center}
		\includegraphics[width=1\linewidth]{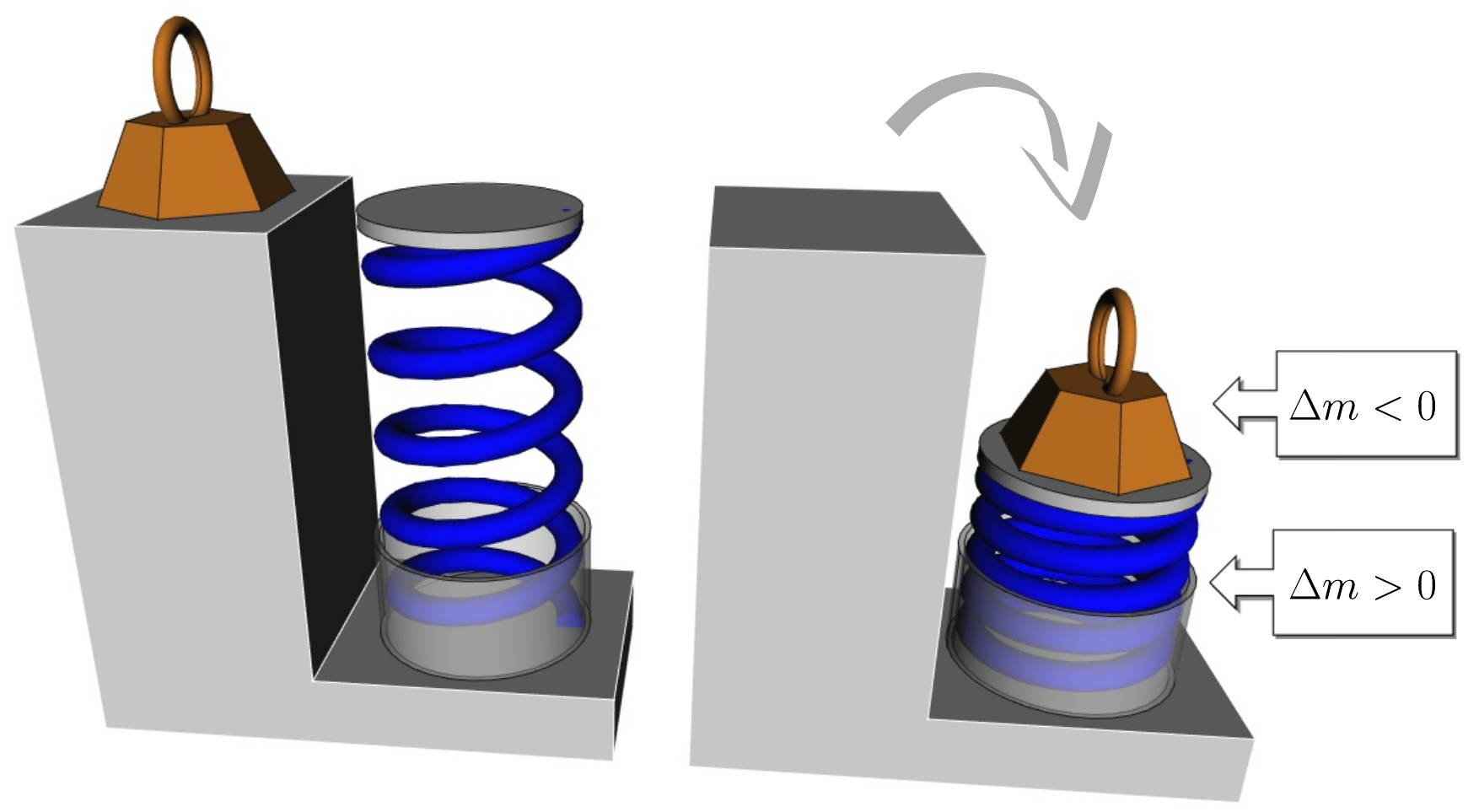}
		\caption{When a weight (in brown) compresses a vertical spring (in blue) it undergoes a decrease of mass (less potential energy), while that of the spring increases (more elastic potential energy).}
		\label{pes}
	\end{center}
\end{figure}

The elastic potential energy of a constrained metal spring is of the same nature (see Fig.\ref{pes}).
\myquote{When work is done on the spring, the spring’s rest energy increases in the form of $\Delta m$}\,\cite{Hecht_2016}.
Compared to the gravitational potential energy of a stone that has an arbitrary zero at the ground level, the spring can be stretched or compressed with an identical restoring force (up to the sign) towards the equilibrium position that unambiguously defines the zero of potential energy of the spring.
This equilibrium position originates from the microscopic net interaction potential between the atoms of the crystal: each atom is in the minimum of a potential well made by the presence of others. The work necessary to constrain the spring is that needed to deviate atoms from this minimum.
The potential energy of the spring is the sum of those of its atoms.

Not only metal springs are elastic. So are pieces of rubber. But contrary to what it was suggested in ref.\,\cite{Hecht_2016} (but this point is marginal in the paper) the origin of this elasticity is different. It has the same origin as that of a volume of gas in thermal equilibrium with its surroundings but at a different pressure.
It is entropic\,\cite{deGennes_1979}.

\subsection{Entropic forces}

Consider a volume of gas in a container equipped with a piston allowing its contents to be compressed or expanded.
Like the spring, the piston has an equilibrium position that corresponds to equal forces applied on it.
When deviating it from this position by pushing or pulling, we feel an elastic restoring force that is apparently comparable to what it would be if the gas were replaced by a spring.
So that it is legitimate to state that when the piston deviates from its equilibrium position the overall system stores an amount of elastic potential energy.
But at a microscopic level, for a perfect gas there is no interaction potential between molecules. Even for real gas, for which pair-interactions can be modelized by a Lennard-Jones potential, interactions can be neglected as soon as the particles are not in contact (between two collisions). Gas particles do not interact at distance and do not have equilibrium position.

\begin{figure}[!htbp]
	\begin{center}
		\includegraphics[width=1\linewidth]{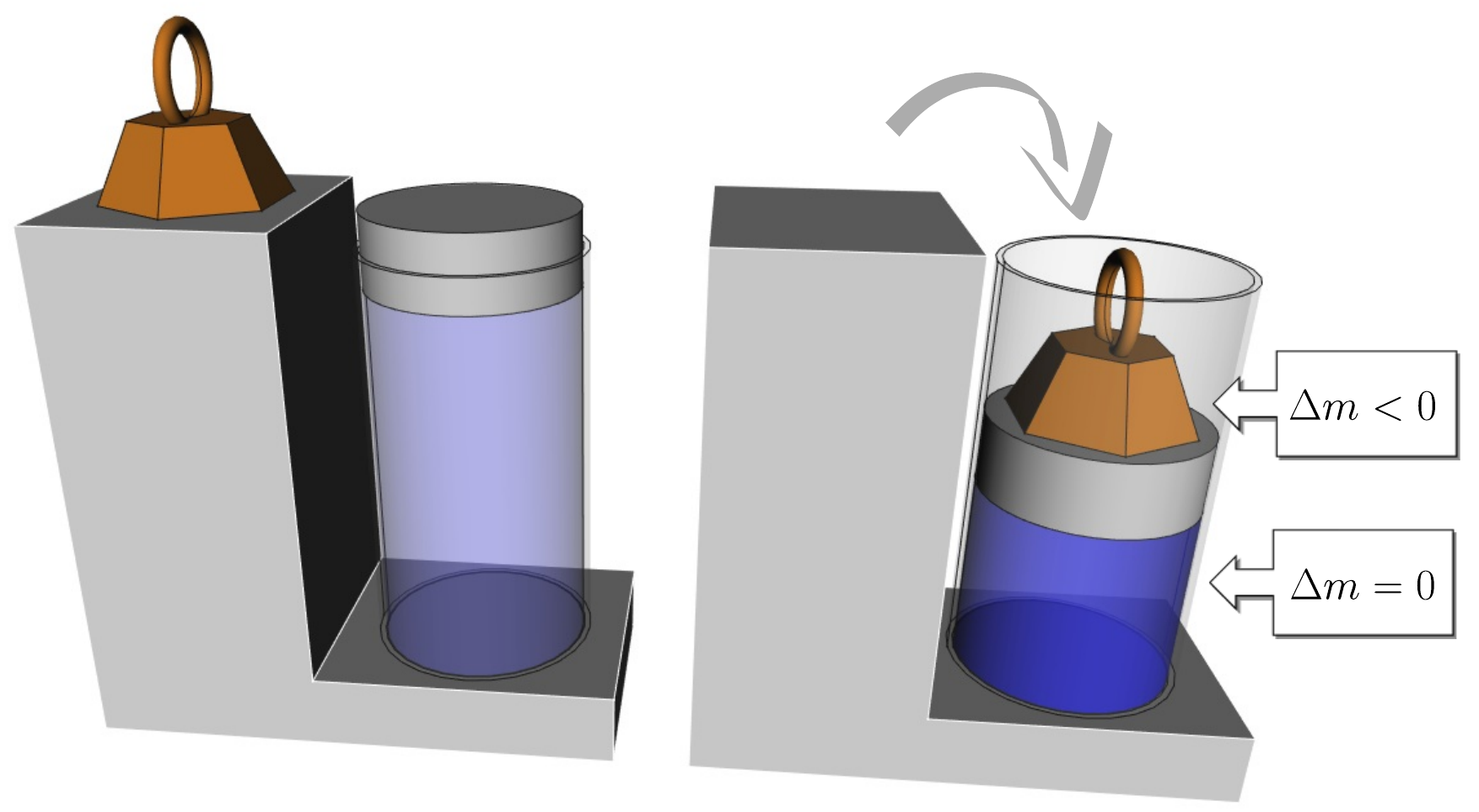}
		\caption{If the spring of Fig.\ref{pes} is replaced by a gas compressed with a piston at constant temperature, the gas has less entropy but the same internal energy and its mass is unchanged.}
		\label{peg}
	\end{center}
\end{figure}

The elastic force we feel on the piston is due to the balance between the many collisions it experiences with the gas molecules on both sides (each collision involving the kinetic energy of one given particle) and the force applied by the surroundings on the piston.

First, note that the internal energy of a system made of $N$ non-interacting independent entities is its temperature $T$ (in Joule).
For the sake of simplicity, let us assume that this is the case for the gas inside the container and for the atmosphere outside (that of the surroundings), which is reasonable in the case where both are air close to atmospheric pressure.
Neglecting interactions means in particular neglecting hydrodynamic interactions, friction and viscosity and thus the time delay to reach the equilibrium after any perturbation.
In terms of thermodynamics it means that the overall system (gas plus surroundings) is always at equilibrium and that the transformation is reversible. The existence of such a reversible process allowing to pass from one state to another is the only way in phenomenological thermodynamics to measure (and thus to define) entropy that is given by its exact differential:
\begin{equation}\label{defS}
	\diff S = \frac{\diff{Q_r}}{T}
\end{equation}
where $Q_r$ is the heat exchanged for a reversible process.

Note that the notion of instantaneous equilibrium, and consequently that of reversible process, appears to be incompatible with special relativity because in principle nothing (and in particular the propagation of a perturbation) can go faster than light. 
The same issue exists for mechanical potential energy\,\cite{Brillouin_1965a, Brillouin_1965b}. But this is not a problem as far as we are concerned by the initial and final states of a process and not by the process in itself (e.g. as far as we consider monothermal and not isothermal processes). The notion of reversible process in thermodynamics is equivalent to the classical mechanics limit of special relativity, which is conditioned on the two assumptions: velocities of particles are small compared to that of light; characteristic distances in the system are small, so are delays in the propagation of signals.

To go further into our problem two cases for the gas container are worth considering:

\begin{enumerate}
	\item the container is adiabatic, i.e. it prevents  heat exchanges with the surroundings; 
	\item the container is diathermal, i.e. it allows heat exchanges with the surrounding.
\end{enumerate}

Consider the gas in an adiabatic container. Compressing the gas by pushing the piston, we produce work and provide to the gas an equivalent amount of energy. Doing so, as the gas cannot dissipate heat, its internal energy necessarily increases, and so its temperature. 
From Eq.\ref{defS}, it follows that if there is no heat exchanged, there is no variation of entropy.
Clearly, the reversible adiabatic case is not the matter of the \textquote{mass-entropy equivalence} that envisages differences of entropy between two states at the same temperature, but it is worth considering to what follows.
Can the increase in internal energy (temperature) be assimilated to the elastic potential energy? No, it cannot. Because when pulling (instead of pushing) the piston, this time the gas decreases in internal energy, but there is still a restoring force and a positive potential energy stored somewhere. The elastic potential energy is in fact stored in the solid container, and in the external mechanical part of the device that drives the piston, under a form quite comparable to that it is for an elastic spring, but it is not stored in the gas.

Consider the ideal diathermal container. Heat exchanges ensure that for any transformation the initial and final states of the gas are  both in thermal equilibrium with the surroundings. If in addition, the surroundings is so large that its temperature can be considered constant, the transformation is monothermal: The temperatures of the initial and final states are the same.
Then by definition, the internal energy of the gas is also unchanged. Whatever the work provided to it, the gas does not store additional energy compared to what it initially contained.
The system can receive work $W$, but it dissipates an equal amount of energy to the surroundings under the form of heat $Q$, or do the reverse. For any monothermal transformation:
\begin{equation}\label{WQ}
	W+Q=0
\end{equation}

The entropy change is given by considering a process slow enough to ensure a constant temperature at all times (isothermal transformation). By integrating Eq.\ref{defS}, one gets
\begin{equation}
	T\Delta S = Q_r = -W_r
\end{equation}
The differential of the work is $\diff W_r=-P\diff V$, with $P$ the pressure and $V$ the volume. As $P=NTV^{-1}$, one has 
$\diff W=-NTV^{-1}\diff V$, integration from $V_0$ to $V$ gives
\begin{equation}\label{defDS}
	T\Delta S = T N \ln (V/V_0)
\end{equation}
So that in the monothermal case, this time under the action of the piston the gas undergoes a change in entropy, but the corresponding elastic potential energy is still not stored by the gas itself. It is stored by the container, the mechanical part that drives the piston and the surroundings.
The gas itself does not store more or less potential (or internal) energy which could correspond to any variation in its mass. In this, it differs from the spring (see Fig.\ref{pes} and \ref{peg}).

The case of a piece of rubber is even more illustrative because there is no need of a container. Rubber is made of cross linked linear polymer chains which form a three dimensional network. Let $N$ be the number of independent chain segments of the chain portion between two first neighbor crosslinks at distance $R_0$ when the rubber is unconstrained. In this state, the chain conformation is random with $R_0$ and $N$ linked by the scaling relation $N\propto R_0^2$.
Stretching the rubber, causes the distance between crosslinks to increase in the same direction, forcing the chain to be less random. The corresponding variation of entropy is such as:
\begin{equation}
	\Delta S \propto -(R/R_0)^2
\end{equation}
But as for the gas, at constant temperature the internal energy is constant (\cite{deGennes_1979} p.31). It follows that the elastic potential energy, even if it originates from the rubber, is not stored inside the rubber but by the mechanical part of the surrounding that is responsible for its stretching.

This result for a gas or a piece of rubber is actually general. The internal energy of a set of independent entities, such as a set of bits, is its temperature. It follows that any monothermal variation of entropy interpreted in terms of potential energy stored under the form of rest mass cannot be localized in such a system, but only in its surroundings.

\section{Logical versus thermodynamical irreversibilities}\label{log_vs_th}

The second law of thermodynamics is two folds. The first is the definition of entropy as a state quantity defined by its exact differential given by Eq.\ref{defS} valid for a reversible process. The second accounts for the general case. It is the Clausius inequality that writes at constant temperature:
\begin{equation}\label{Clausius_ineq}
	-Q\ge -T\Delta S
\end{equation}
This means that the quantity $-Q$ of heat dissipated (and received by the surroundings) is always higher than  $-T\Delta S$.
For a system made of independent entities at constant temperature, the internal energy is also constant (Eq.\ref{WQ}) so that the heat dissipation is compensated by the same amount of work ($W=-Q$) provided to the system. Generally, heat is unwanted and work is more valued and can be viewed as an energy cost. With Eq.\ref{Shannon1} the Clausius inequality rewrites in this case:
\begin{equation}\label{cineq2}
	W\ge - T\Delta H \ln 2
\end{equation}
Consider the process of reducing the volume of the phase space of the dynamical system ($\Delta S\le 0$). The uncertainty about the current microstate of the system, or the quantity of information it emits, is reduced by $\Delta S/\ln 2$. The total amount of information we lack to describe the system is reduced accordingly, as if we had acquired $\Delta H=\Delta S/\ln 2$ bits of data about the system.
So that Eq.\ref{cineq2} can be expressed per bit ($\Delta H=-1$) of acquired data:
\begin{equation}\label{Wbit1}
W_\text{acq/bit} \ge T \ln 2
\end{equation}
which expresses that $T\ln 2$ is the minimum energy cost to acquire 1 bit of data about the dynamical system under consideration.
This statement is nothing more than a reformulation of the Clausius inequality in terms of quantity of information.

\begin{figure}[!htbp]
	\begin{center}
		\includegraphics[width=1\linewidth]{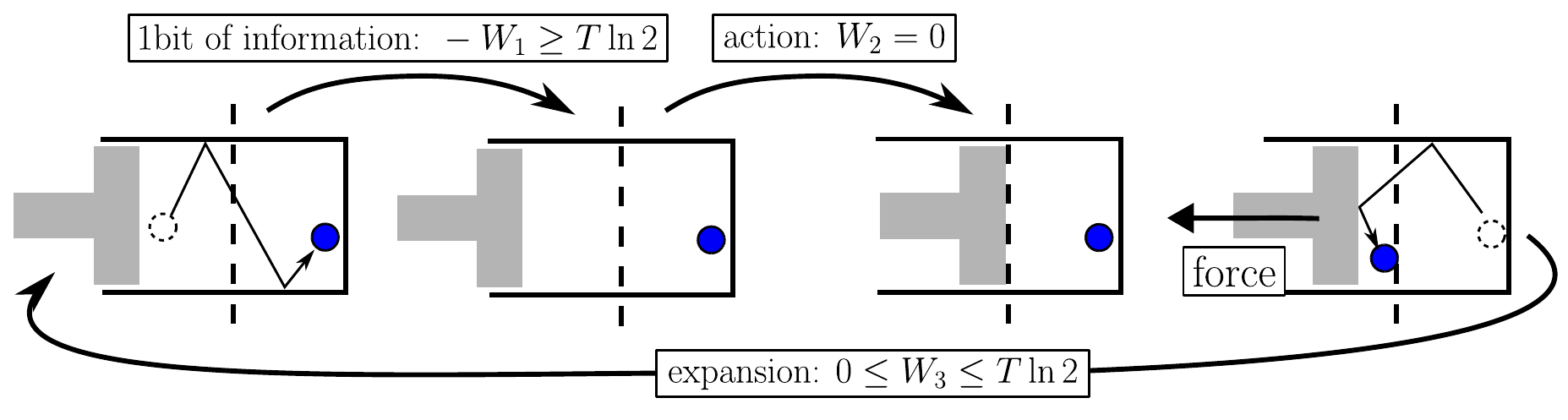}
	\caption{Szilard demon detects when the particle (in blue) is in the suitable compartment. In doing so, it acquires 1 bit of information for an energy cost $-W_1\ge T\ln 2$ (Eq.\ref{Wbit1}). Then it can install a piston for free ($W_2=0$), allowing the device to subsequently produce work ($W_3\le T\ln 2$). The overall energy balance is $W_3+W_1+W_2\le 0$.}		
	\label{szilard}
	\end{center}
\end{figure}

Equation \ref{Wbit1} suffices in itself to understand the functioning of demonic engines like that of Maxwell\,\cite{Maxwell_1872} or the simplified version of Szilard\,\cite{Szilard_1964} (see Fig.\ref{szilard}) and their physical implementations under the form of ratchet-pawl mechanisms\,\cite{Feynmann_Ratchet, Brillouin1950}.
Each bit of information needed for the engine to work has a minimum energy cost of $T\ln 2$. This is a direct application of the second law that provides the link between information and energy.

The Landauer principle\,\cite{Landauer_1961, Bennett_1982, Bennett_2003} reaches the same result without information theory but with an indirect application of the second law.
It is at the heart of the \textquote{mass-energy-information equivalence principle}\,\cite{Vopson_2019, Vopson_2022}. Let us report some quotes:

\myquote{The M/E/I principle [the mass-energy-information equivalence principle] states that information is a form of matter, it is physical, and it can be identified by a specific mass per bit while it stores information or by an energy dissipation following the irreversible information erasure operation, as dictated by the Landauer principle.} (Vopson\,\cite{Vopson_2022})

\myquote{We demonstrated that the Landauer principle supplies the estimation of the minimal mass of the particle allowing the recording/erasure of information within the surrounding medium at temperature $T$} (Bormashenko\,\cite{Bormashenko_2019b}).

The purpose of this section is not to discuss the reasonings leading to these conclusions, but rather to show that the root of them, i.e. the Landauer principle, is not correct because it results: 1)~from considering a doubly particular case; 2)~from a confusion between logical and thermodynamical irreversibilities.

\subsection{Landauer \textquote{principle} is a doubly particular case}\label{secLandauer}

At the basis of the Landauer principle allowing Eq.\ref{Wbit1} to be found without any reference to the Shannon's encoding problem are the two assumptions below:
\begin{enumerate}
	\item For a cyclic process (such as that of the Szilard engine), the recording or acquisition of a data bit is supposed to first require the erasure of that bit.
	\item The erasure of a data bit is supposed to necessarily involve an irreversible non-quasistatic stage (i.e. uncontrollable and similar to the free expansion of a gas), so that when performed cyclically it has a minimum energy cost of $T\ln 2$.
\end{enumerate}
The first supposed requirement will be discussed in the second part of this section. Here, we only focus on the second.

Landauer and Bennett first imagine a one-to-one correspondence between logic and thermodynamic states. They imagine a particle in a bi-stable potential well allowing two different positions (labeled 0 and 1, respectively) to be equally stable. The ERASE operation consists is putting the particle in position 0 (SET TO 0). It is done in three elementary stages or operations.

\vspace{0.25cm}
\noindent
Landauer's ERASE procedure:
\begin{enumerate}
	\item SET TO S (standard state): set the particle in an undetermined position by lowering the energy barrier between the two positions.
	\item BIAS TO 0: apply a bias favoring the zero position.
	\item STABILIZE: raise the energy barrier and stop the bias.
\end{enumerate}
The point is that the path chosen by Landauer to achieve the first stage, throws the probability density of the particle position out of control and causes it to leak from one potential well to the other. It is similar to the free expansion of a gas, initially confined in one half-volume of a box (with label 0 or 1) and suddenly allowed to occupy the entire volume. Whereas the last two stages can be done in a quasistatic manner equivalent to the isothermal compression the gas in the correct half-volume of the box (with label 0). During the first stage, neither heat nor work are exchanged with the surroundings, whereas the last two have an energy cost at least equal to $T\ln 2$. The net energy balance of the total operation is thus:
\begin{equation}\label{Landauer}
	W_\text{erase 1 bit} \ge T \ln 2
\end{equation}
Conjointly with the necessity to erase prior to acquire 1~bit of data, this last equation allows us to found Eq.\ref{Wbit1}.

The non-quasistatic irreversibility of the first stage is supposed by Landauer and Bennett to be unavoidable. In a previous paper\,\cite{Lairez_2023}, it has been shown that the best way to avoid any probability-density leakage between the two potential-wells is to have only one, but still two logical states. This is subject to the possibility of establishing a two-to-one correspondence between logic and thermodynamic states. An example of such correspondence has been given in ref.\,\cite{Lairez_2023} which ruins the generality of the Landauer principle. For the purpose of this paper, let us give another counter-example.

\begin{figure}[!htbp]
	\begin{center}
		\includegraphics[width=1\linewidth]{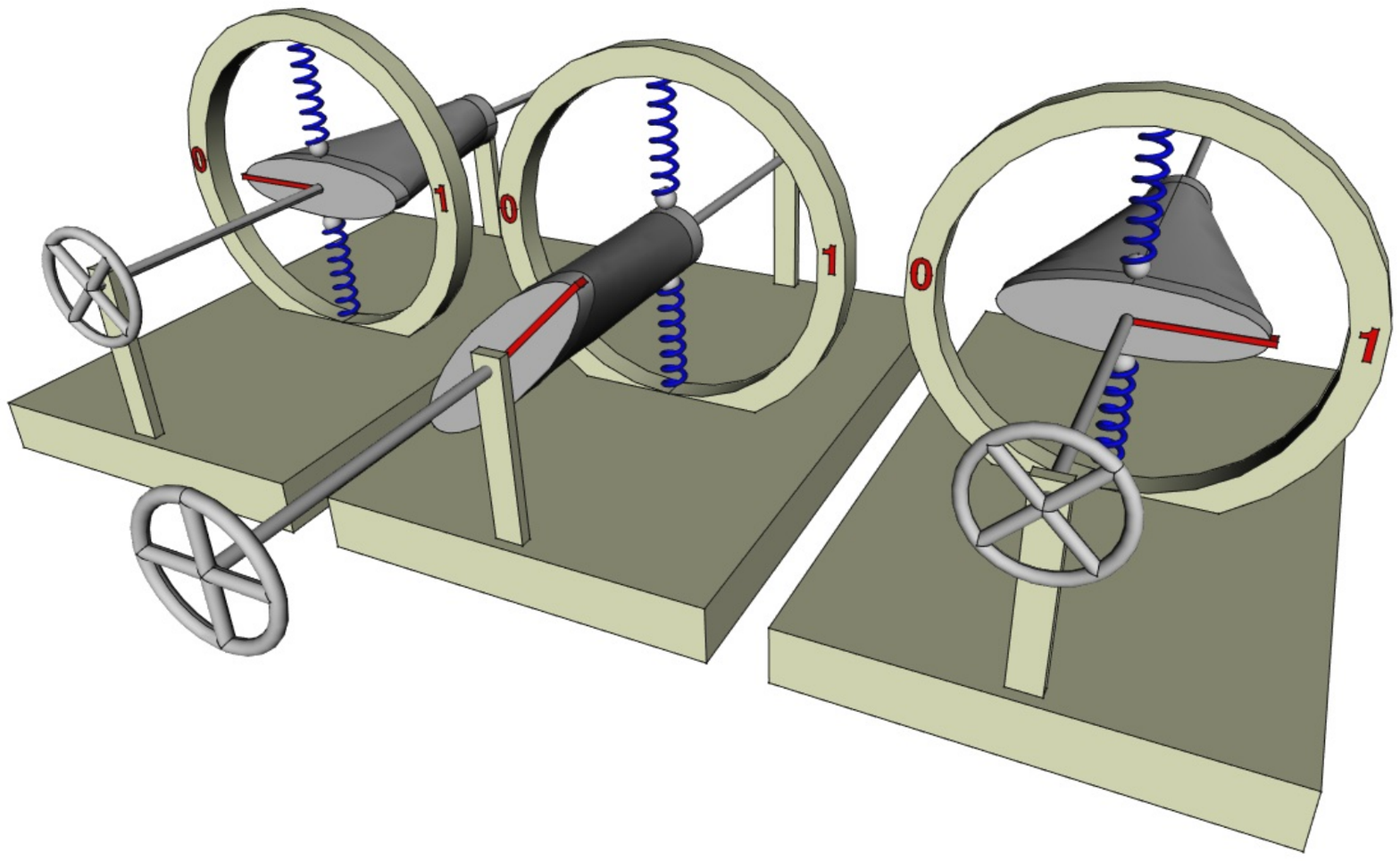}
		\caption{Two-to-one correspondence between logic and thermodynamic states. The bit-state (in red) is represented by the angle of a cam centered on a rotating axis controlled by a steering wheel. The cam-profile is elliptical in front and circular in back and can constrain two vertical springs (in blue) that define the thermodynamic (or mechanical) state. When the steering wheel is pushed (left and right) the bit has two stable logical states 0 or 1. By slowly pulling the steering wheel, the energy barrier between the two states vanishes in a quasistatic manner and drives the bit (middle) in an undetermined S-state (following Landauer-Bennet terminology).}
		\label{mono}
	\end{center}
\end{figure}

In Fig.\ref{mono}, we imagine a cam that can compress two springs (which can be replaced if we want by two volumes of gas compressed by pistons). The angle of the cam defines the bit state, whereas the state of the springs defines the thermodynamic (or mechanical) state. The cam has a smooth shape that continuously passes from an elliptical section on one side (front) to a circular one on the other (rear), both being centered on the axis of rotation provided with a steering wheel allowing it to be driven. When the steering wheel is pushed (left and right in Fig.\ref{mono}), the bit is stabilized in position 0 or 1 by the two springs. In both positions, the constraint they undergo are the same and it is not possible to know the bit-state by simply observing the state of the springs. There is actually only one thermodynamic state and thus a two-to-one correspondence between logic and thermodynamic states. The ERASE operation can be done by pulling the steering wheel, so that the springs are in contact with the circular section of the cam and the energy barrier between the two logical states is zero, the bit is then in the undetermined S-state.
Then, turn the steering wheel in order to align the red mark of the cam with position 0 (apply a bias), finally push the steering wheel to restore the energy barrier. The entire operation only involves friction, so that the heat dissipation tends to zero as the velocity of the steering wheel manipulation tends to zero. It is quasistatic. Note that if in addition, the small radius of the ellipse is equal to the radius of the circle (as in Fig.\ref{mono}), the state of constraint of the springs is the same whether the bit is in position 0, D  or 1. So that the entire ERASE operation can be done at constant elastic potential energy of the springs. Note also that in the two logic states (0 and 1), as the elastic potential energies of the springs are the same, their mass in the framework of special relativity is also the same. A set of independent bits, built with this physical implementation, can be erased in a quasistatic manner and with no variation of rest mass.

\begin{figure}[!htbp]
	\begin{center}
		\includegraphics[width=1\linewidth]{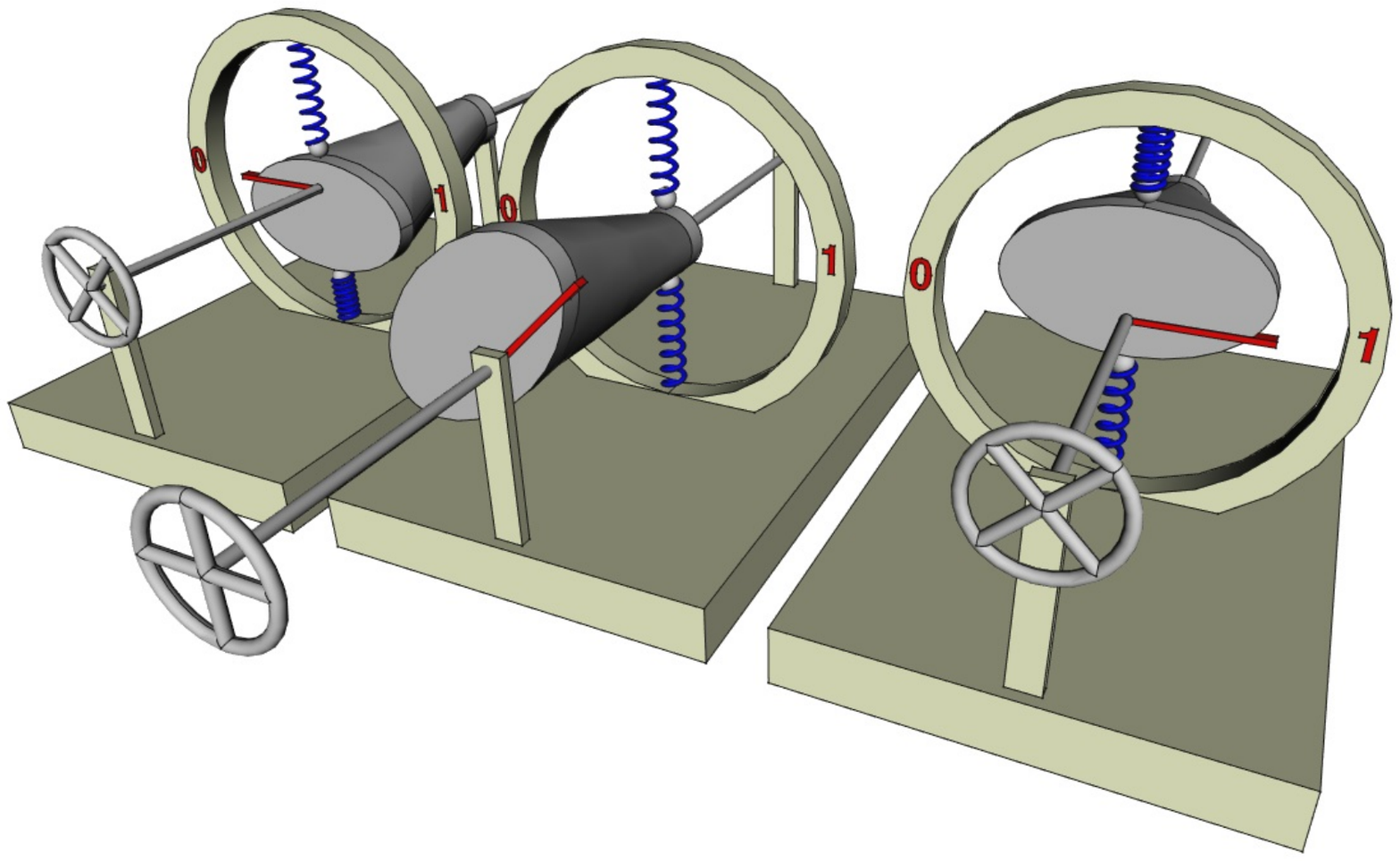}
		\caption{One-to-one correspondence between logic and thermodynamic states. The mechanism is similar to that of Fig.\ref{mono} but this time the cam is not centered in its elliptical part (but is still centered in its circular part). So that depending on the bit state (in red), the two springs are not equally constrained allowing to identify two different thermodynamic (or mechanical) states (in blue) each one corresponding to a different logical state. Just like for the two-to-one implementation (Fig.\ref{mono}) the bit can be set to 0 (erased) in a quasistatic manner that only involves friction and avoids any leakage between the two states.}
		\label{bi}
	\end{center}
\end{figure}

A two-to-one correspondence could be viewed as a very particular case of physical implementation of logic and the one-to-one correspondence perceived as the general case. Interestingly, the example just given above can be slightly modified in order to obtain a one-to-one correspondence similar to that of Landauer but with no leak. For this, it is enough to decentre the elliptical face of the cam (while the circular face remains centered) as it is shown in Fig.\ref{bi}. Then, the logical states 0 or 1 are still stable and well separated by an energy barrier, but the state of the springs are not the same in both cases. It is now possible to know the bit-state by only observing in which state are the springs (compressed up or down). There is a one-to-one correspondence between logic and thermodynamic states. The procedure to change the bit state (e.g. SET TO 0 or ERASE) is the same as in the previous case and is quasistatic. During this operation, the potential energy of the springs is first released (when pulling the steering wheel), then the same amount is stored again (when pushing the steering wheel).
The sum of the potential energy of the springs is the same whatever the bit state so that it can be set to 0 with no change in rest mass.

From these two counter-examples, it appears clearly that logical and thermodynamical irreversibilities are not linked. The reason for this is explained just below.

\subsection{Irreversibilities}

The logical irreversibility of an operation is defined by Landauer: \myquote{We shall call a device [an operation] logically irreversible if the output of a device does not uniquely define the inputs.} (Landauer \cite{Landauer_1961}). 
The ERASE operation is logically irreversible: two possible initial states 0 or 1 (the input of the operation) lead to only one final state 0 (the output). 
Further in the same paper, Landauer writes: \myquote{Logical irreversibility, we believe, in turn implies physical [thermodynamical] irreversibility}. This last point is discussed in this section.

As soon as we deal with the physical implementation of a logical operation on a bit, this operation becomes a thermodynamical transformation (or a process).
With the above definition, it is clear that the logical irreversibility of an operation is defined only by the properties of the bit before and after the transformation. The logical irreversibility is a property of the initial and final states of the bit. It is not a property of the path that has been used to perform the transformation.

In other words, a transformation (say a transport) from $A$ to $B$ is logically irreversible if once in $B$ the information from where the system started has been lost, so that it is not possible to return to the starting point $A$. The same transformation is thermodynamically irreversible if it is not possible to return to $A$ by using the same path backward.
Thermodynamical irreversibility is a path property.

Due to this fundamental difference, logical and thermodynamical irreversibilities cannot be linked by a material implication.

In thermodynamics, when we deal with the path allowing a system to change from an initial to a final state, we first wonder whether or not it can be decomposed in a succession of infinitely small changes, i.e. in a succession of quasi-equilibrium states. In other words, consider the variation of entropy versus the extent of the change, we wonder whether or not this variation is differentiable. If so, the path is said quasistatic and can be potentially reversible if slow down enough. Otherwise, the very point where the discontinuity occurs is an inherently irreversible step.
At this point, the process evolves spontaneously in an uncontrollable manner. This occurs when a system suddenly finds itself far from equilibrium when an internal constraint has been released (a typical example is that of the free expansion of a gas).
Consider such an irreducible step as a process in itself. No work can be extracted from it ($W=0$). At a given temperature, according to Eq.\ref{WQ} and \ref{Clausius_ineq}, one gets:
\begin{equation}
	\Delta S>0
\end{equation}
An increase of entropy is in fact a necessary condition for an inherently irreversible (non-quasistatic and irreducible) process to occur. But it is not at all a sufficient condition. The same change in entropy can occur using a reversible path, otherwise it would not be defined in thermodynamics because it could not be associated with any measurable quantity.

Consider the ERASE operation on the thermostatistics side and let $\Gamma$ be the phase space. 
The two initial possibilities $\Gamma=\{0, 1\}$ result in only one $\Gamma=\{0\}$ once the bit has been erased. So that 
\begin{equation}
\Delta S_\text{ERASE}<0	
\end{equation}
This means that ERASE is not inherently irreversible, but that
any thermodynamic path leading to this operation can be decomposed into elementary steps (just as Landauer did).
In fact, in Landauer's physical implementation, only the first step (the one which brings the system into an indeterminate state S) can possibly be intrinsically irreversible, because  $\Gamma$ initially $\{0, 1\}$ changes into  $[0,1]$, so that
\begin{equation}
	\Delta S_\text{SET TO D}>0	
\end{equation}
For this step, another path than that of Landauer can be chosen, quasistatic this time, as shown in section \ref{secLandauer}.

\myquote{[Logical irreversible] operations are quite numerous in computer programs as ordinary written: besides erasure, they include overwriting of data by other data} (Bennett\,\cite{Bennett_1982}). 
On the thermostatics side, the status of OVERWRITE depends on which new data replaces the old ones. For cyclic recording of data, as that needed when implementing a Szilard engine, if the system is in a stationary regime, the probability distribution of the data is unchanged from one cycle to the next. So that  overwriting old data by new ones leaves the phase space unchanged, thus:
\begin{equation}
	\Delta S_\text{OVERWRITE}=0
\end{equation}
It follows that OVERWRITE is also not an inherent irreversible process. It can be decomposed in quasistatic steps. To \textquote{elucidate} the functioning of the Szilard engine, Landauer-Bennett chose to break down OVERWRITE into ERASE then WRITE, which brings us back to the previous discussion about ERASE. But OVERWRITE can be done in a direct manner without ERASE, just like an old magnetic tape can be reused without to be reset in a virgin state (erased).

Logical irreversibility is not linked to thermodynamic irreversibility.
There is no conceptual impediment for a logical irreversible operation to be quasistatic and for heat dissipation to vanish as the operation slow down.

\section{Information is dynamical}\label{inf_dyn}

\myquote{To test the hypothesis [the mass-energy-information equivalence principle] we propose here an experiment, predicting that the mass of a data storage device would increase by a small amount when is full of digital information relative to its mass in erased state. For 1Tb device the estimated mass change is $2.5 \times 10^{-25}$\,Kg.} (Vopson\,\cite{Vopson_2019}). 
Beyond the already recognized difficulty of carrying out such a measurement\,\cite{Vopson_2022}, we will show here that this idea is nonsense and inconsistent with everything we know about thermodynamics (and physics).

The first argument directly comes from the fundamental difference between logical and thermodynamical irreversibilities that has been exposed in the previous section.
Consider the above hard drive full of data
and the three experiments below: 
\begin{enumerate}
	\item Directly erase the hard drive. This operation is logical irreversible (it is impossible to retrieve the data once they have been erased).
	\item Make a copy of the hard drive then erase the original one. This operation is logical reversible (with the copy it is possible to restore the original hard drive to its original state).
	\item Make a copy, erase one hard drive then the second. The erasure of the first hard drive is logical reversible whereas that of the second is irreversible.
\end{enumerate}
For the copy to be of any use in preserving data integrity, it must be physically independent from the original (we can imagine moving it to the other side of the earth). It follows that the mass defect (if there is one) that would be measured for the above four erase operations would have exactly the same value. 
If there is a mass defect, it has nothing to do with logical irreversibility nor with information that would be lost or not.

The independence of two hard drive also holds for two bits. This is implicit in the mass-energy-information \textquote{equivalence} when it is expressed per bit: \myquote{Using the mass-energy-information equivalence principle, the rest mass of a digital bit of information at room temperature is $m_{bit}=3.19\times 10^{-38}$\,kg.} (Vopson\,\cite{Vopson_2020}). But sometime it is explicit: \myquote{Essentially, a bit of information could be seen as an abstract information particle, with no charge, no spin, and rest mass} (Vopson\,\cite{Vopson_2020}).
It follows that the above three experiments performed with a hard drive could be done with a bit with the same conclusion. 

Data are stored with bits set either to 0 or 1, the two values equally participate to the storage of information. If a bit of information has a rest mass, the latter is independent of its value 0 or 1. The ERASE operation is usually presented as a SET TO 0 operation, but this is a convention and it could be SET TO 1 (as Landauer did in his first paper \cite{Landauer_1961}). This suggest another experiment in two stages:
\begin{enumerate}
	\item Erase (SET TO 0) one given bit of information (with value either 0 or 1).
	\item Erase it a second time.
\end{enumerate}
According to the mass-energy-information equivalence principle, a mass defect should be observed in the first stage, while it should not be observed in the second. The only explanation for this difference should be the change in entropy caused by the ERASE: in the first stage $\Gamma: \{0, 1\}\rightarrow \{0\}$, whereas for the second stage  $\Gamma: \{0\}\rightarrow \{0\}$. 
This brings us back to the first section of this paper: at constant temperature no change in rest mass accompanies a change in entropy.

In fact, the data stored in the hard drives or the bits above are not information in the sense given to that word by Shannon. The physical support of these data can be considered as a thermodynamic system in its own right. But for this, it must be read and emit outcomes just like other dynamical systems do.
To consider these data as information, they must not be detached from their subject matter, i.e. from the dynamical system that emits this information. Once detached from this dynamical system, the information becomes frozen and outdated, it has no value and no link with energy. Let us detail this point.

In the word \textquote{thermodynamic} there is \textquote{dynamic}. This is a truism but apparently worth remembering: $\Delta S$ must be multiplied by temperature $T$ to become an energy. In other words, the link between energy and information only exists when the renewal of the latter obeys to the same dynamics as that of the system it concerns.
This is particularly clear with the Szilard engine. Imagine that the position of the particle is recorded on a hard drive for a given time interval. Once this is done, these old data cannot be of any utility to extract  energy from the current system in return to that spent on their acquisition and recording.

\section*{Conclusion}

Entropy (and quantity of information) is a concept.
Just like potential energy is.
There are actually many common points between them.
For instance, just like potential energy, it is not possible to measure entropy without changing it into something else (i.e. changing potential energy into kinetic energy and changing entropy into heat). Also, zeros for both quantities may appear arbitrary and not intrinsic to the nature of things. Nevertheless, these concepts have some fundamental differences. Potential energy was introduced to satisfy a conservation principle for energy (first law of thermodynamics), while entropy was introduced to account for the irreversibility of changes in the form of energy (second law), that is to say a change in quality but not in quantity. Basically, this difference rules out the idea of a mass-entropy equivalence (section \ref{pe_vs_e}).

This idea of a mass-entropy equivalence (or mass-information equivalence) is actually the last attempt to materialize the link between information and energy, that is to say to make it more \textquote{physical}, more tangible, less elusive. It originates from the Landauer principle. The latter is actually due to a confusion between logical irreversibility (that is a state dependent property) and thermodynamical irreversibility (that is a path dependent property). Once this has been clarified, it appears that there is no finite limit of heat dissipation to erase a bit. In other words, a bit does not store more energy whether it is set to a given data value or erased (section \ref{log_vs_th}).

This brings us to the last confusion at the origin of the mass-information equivalence: stored data is not information in the Shannon sense. Stored data are frozen, information is dynamical. Stored data are actually outcomes of a dynamical system that have been acquired (thus brought to our knowledge) then copied somewhere (stored). But information is very special, as soon it has been given (acquired), it is no longer information, it is outdated. Information cannot be given twice. The link between energy and information is that of a dynamical system as a source of information in the spirit of Shannon
(section \ref{inf_dyn}). Just like the link between energy and entropy.

In 1948, when Shannon\,\cite{Shannon_1948} identified the minimum number of bits (which he called quantity of information) to encode the behavior of a dynamic system as its statistical entropy, this was a great advance: entropy became information.
Although this alone was of great importance for computer scientists or for communications engineers, for physicists the major breakthrough did not lie in this identification, which may appear to them as a simple question of vocabulary. The major breakthrough  was in the second part of work of Shannon who identified also this quantity of information with a measure of the uncertainty about the system. The resulting principle of maximum entropy\,\cite{Jaynes_1957, Jaynes_1968, Shore_1980} made it possible to legitimize \textit{a priori} probabilities and resolve many inconsistencies in statistical mechanics (for a review on this topic see \cite{Lairez_2023c}).

Entropy is information, as fascinating as that may be, we must not forget that the relationship also holds in the other direction: information is entropy and is just that.

\bibliography{\string~/Documents/Articles/weri_biblio.bib}

\begin{thebibliography}{26}%
\makeatletter
\providecommand \@ifxundefined [1]{%
 \@ifx{#1\undefined}
}%
\providecommand \@ifnum [1]{%
 \ifnum #1\expandafter \@firstoftwo
 \else \expandafter \@secondoftwo
 \fi
}%
\providecommand \@ifx [1]{%
 \ifx #1\expandafter \@firstoftwo
 \else \expandafter \@secondoftwo
 \fi
}%
\providecommand \natexlab [1]{#1}%
\providecommand \enquote  [1]{``#1''}%
\providecommand \bibnamefont  [1]{#1}%
\providecommand \bibfnamefont [1]{#1}%
\providecommand \citenamefont [1]{#1}%
\providecommand \href@noop [0]{\@secondoftwo}%
\providecommand \href [0]{\begingroup \@sanitize@url \@href}%
\providecommand \@href[1]{\@@startlink{#1}\@@href}%
\providecommand \@@href[1]{\endgroup#1\@@endlink}%
\providecommand \@sanitize@url [0]{\catcode `\\12\catcode `\$12\catcode
  `\&12\catcode `\#12\catcode `\^12\catcode `\_12\catcode `\%12\relax}%
\providecommand \@@startlink[1]{}%
\providecommand \@@endlink[0]{}%
\providecommand \url  [0]{\begingroup\@sanitize@url \@url }%
\providecommand \@url [1]{\endgroup\@href {#1}{\urlprefix }}%
\providecommand \urlprefix  [0]{URL }%
\providecommand \Eprint [0]{\href }%
\providecommand \doibase [0]{https://doi.org/}%
\providecommand \selectlanguage [0]{\@gobble}%
\providecommand \bibinfo  [0]{\@secondoftwo}%
\providecommand \bibfield  [0]{\@secondoftwo}%
\providecommand \translation [1]{[#1]}%
\providecommand \BibitemOpen [0]{}%
\providecommand \bibitemStop [0]{}%
\providecommand \bibitemNoStop [0]{.\EOS\space}%
\providecommand \EOS [0]{\spacefactor3000\relax}%
\providecommand \BibitemShut  [1]{\csname bibitem#1\endcsname}%
\let\auto@bib@innerbib\@empty
\bibitem [{\citenamefont {Maxwell}(1872)}]{Maxwell_1872}%
  \BibitemOpen
  \bibfield  {author} {\bibinfo {author} {\bibfnamefont {J.~C.}\ \bibnamefont
  {Maxwell}},\ }\href
  {https://books.google.fr/books?id=5u84AAAAMAAJ&printsec=frontcover&hl=fr#v=onepage&q&f=false}
  {\emph {\bibinfo {title} {Theory of heat}}},\ \bibinfo {edition} {3rd}\ ed.\
  (\bibinfo  {publisher} {Longmans, Green and Co.},\ \bibinfo {address}
  {London},\ \bibinfo {year} {1872})\BibitemShut {NoStop}%
\bibitem [{\citenamefont {Shannon}(1948)}]{Shannon_1948}%
  \BibitemOpen
  \bibfield  {author} {\bibinfo {author} {\bibfnamefont {C.~E.}\ \bibnamefont
  {Shannon}},\ }\bibfield  {title} {\bibinfo {title} {A mathematical theory of
  communication},\ }\href {https://doi.org/10.1002/j.1538-7305.1948.tb01338.x}
  {\bibfield  {journal} {\bibinfo  {journal} {The Bell System Technical
  Journal}\ }\textbf {\bibinfo {volume} {27}},\ \bibinfo {pages} {379}
  (\bibinfo {year} {1948})}\BibitemShut {NoStop}%
\bibitem [{\citenamefont {Landauer}(1961)}]{Landauer_1961}%
  \BibitemOpen
  \bibfield  {author} {\bibinfo {author} {\bibfnamefont {R.}~\bibnamefont
  {Landauer}},\ }\bibfield  {title} {\bibinfo {title} {Irreversibility and heat
  generation in the computing process},\ }\href
  {https://doi.org/10.1147/rd.53.0183} {\bibfield  {journal} {\bibinfo
  {journal} {{IBM} Journal of Research and Development}\ }\textbf {\bibinfo
  {volume} {5}},\ \bibinfo {pages} {183} (\bibinfo {year} {1961})}\BibitemShut
  {NoStop}%
\bibitem [{\citenamefont {Bennett}(1982)}]{Bennett_1982}%
  \BibitemOpen
  \bibfield  {author} {\bibinfo {author} {\bibfnamefont {C.~H.}\ \bibnamefont
  {Bennett}},\ }\bibfield  {title} {\bibinfo {title} {The thermodynamics of
  computation - a review},\ }\href {https://doi.org/10.1007/bf02084158}
  {\bibfield  {journal} {\bibinfo  {journal} {International Journal of
  Theoretical Physics}\ }\textbf {\bibinfo {volume} {21}},\ \bibinfo {pages}
  {905} (\bibinfo {year} {1982})}\BibitemShut {NoStop}%
\bibitem [{\citenamefont {Bennett}(2003)}]{Bennett_2003}%
  \BibitemOpen
  \bibfield  {author} {\bibinfo {author} {\bibfnamefont {C.~H.}\ \bibnamefont
  {Bennett}},\ }\bibfield  {title} {\bibinfo {title} {Notes on
  {L}andauer{\textquotesingle}s principle, reversible computation, and
  {M}axwell{\textquotesingle}s demon},\ }\href
  {https://doi.org/10.1016/s1355-2198(03)00039-x} {\bibfield  {journal}
  {\bibinfo  {journal} {Studies in History and Philosophy of Science Part B:
  Studies in History and Philosophy of Modern Physics}\ }\textbf {\bibinfo
  {volume} {34}},\ \bibinfo {pages} {501} (\bibinfo {year} {2003})}\BibitemShut
  {NoStop}%
\bibitem [{\citenamefont {Landauer}(1991)}]{Landauer_1991}%
  \BibitemOpen
  \bibfield  {author} {\bibinfo {author} {\bibfnamefont {R.}~\bibnamefont
  {Landauer}},\ }\bibfield  {title} {\bibinfo {title} {Information is
  physical},\ }\href {https://doi.org/10.1063/1.881299} {\bibfield  {journal}
  {\bibinfo  {journal} {Physics Today}\ }\textbf {\bibinfo {volume} {44}},\
  \bibinfo {pages} {23} (\bibinfo {year} {1991})}\BibitemShut {NoStop}%
\bibitem [{\citenamefont {Vopson}(2019)}]{Vopson_2019}%
  \BibitemOpen
  \bibfield  {author} {\bibinfo {author} {\bibfnamefont {M.~M.}\ \bibnamefont
  {Vopson}},\ }\bibfield  {title} {\bibinfo {title} {The
  mass-energy-information equivalence principle},\ }\href
  {https://doi.org/10.1063/1.5123794} {\bibfield  {journal} {\bibinfo
  {journal} {{AIP} Advances}\ }\textbf {\bibinfo {volume} {9}},\ \bibinfo
  {pages} {095206} (\bibinfo {year} {2019})}\BibitemShut {NoStop}%
\bibitem [{\citenamefont {Vopson}(2020)}]{Vopson_2020}%
  \BibitemOpen
  \bibfield  {author} {\bibinfo {author} {\bibfnamefont {M.~M.}\ \bibnamefont
  {Vopson}},\ }\bibfield  {title} {\bibinfo {title} {The information
  catastrophe},\ }\bibfield  {journal} {\bibinfo  {journal} {AIP Advances}\
  }\textbf {\bibinfo {volume} {10}},\ \href {https://doi.org/10.1063/5.0019941}
  {10.1063/5.0019941} (\bibinfo {year} {2020})\BibitemShut {NoStop}%
\bibitem [{\citenamefont {Vopson}(2022)}]{Vopson_2022}%
  \BibitemOpen
  \bibfield  {author} {\bibinfo {author} {\bibfnamefont {M.~M.}\ \bibnamefont
  {Vopson}},\ }\bibfield  {title} {\bibinfo {title} {Experimental protocol for
  testing the mass-energy-information equivalence principle},\ }\href
  {https://doi.org/10.1063/5.0087175} {\bibfield  {journal} {\bibinfo
  {journal} {{AIP} Advances}\ }\textbf {\bibinfo {volume} {12}},\ \bibinfo
  {pages} {035311} (\bibinfo {year} {2022})}\BibitemShut {NoStop}%
\bibitem [{\citenamefont {D\v{z}aferovi\'c-Ma\v{s}i\'c}(2021)}]{Masic_2021}%
  \BibitemOpen
  \bibfield  {author} {\bibinfo {author} {\bibfnamefont {E.}~\bibnamefont
  {D\v{z}aferovi\'c-Ma\v{s}i\'c}},\ }\bibfield  {title} {\bibinfo {title}
  {Missing information in the universe as a dark matter candidate based on the
  mass-energy-information equivalence principle},\ }\href
  {https://doi.org/10.1088/1742-6596/1814/1/012006} {\bibfield  {journal}
  {\bibinfo  {journal} {Journal of Physics: Conference Series}\ }\textbf
  {\bibinfo {volume} {1814}},\ \bibinfo {pages} {012006} (\bibinfo {year}
  {2021})}\BibitemShut {NoStop}%
\bibitem [{\citenamefont {Burgin}\ and\ \citenamefont
  {Mikkilineni}(2022)}]{Burgin_2022}%
  \BibitemOpen
  \bibfield  {author} {\bibinfo {author} {\bibfnamefont {M.}~\bibnamefont
  {Burgin}}\ and\ \bibinfo {author} {\bibfnamefont {R.}~\bibnamefont
  {Mikkilineni}},\ }\bibfield  {title} {\bibinfo {title} {Is information
  physical and does it have mass?},\ }\href
  {https://doi.org/10.3390/info13110540} {\bibfield  {journal} {\bibinfo
  {journal} {Information}\ }\textbf {\bibinfo {volume} {13}},\ \bibinfo {pages}
  {540} (\bibinfo {year} {2022})}\BibitemShut {NoStop}%
\bibitem [{\citenamefont {Lairez}(2023{\natexlab{a}})}]{Lairez_2023}%
  \BibitemOpen
  \bibfield  {author} {\bibinfo {author} {\bibfnamefont {D.}~\bibnamefont
  {Lairez}},\ }\bibfield  {title} {\bibinfo {title} {Thermodynamical versus
  logical irreversibility: A concrete objection to {L}andauer's principle},\
  }\href {https://doi.org/10.3390/e25081155} {\bibfield  {journal} {\bibinfo
  {journal} {Entropy}\ }\textbf {\bibinfo {volume} {25}},\ \bibinfo {pages}
  {1155} (\bibinfo {year} {2023}{\natexlab{a}})}\BibitemShut {NoStop}%
\bibitem [{\citenamefont {Clausius}(1879)}]{Clausius_1879}%
  \BibitemOpen
  \bibfield  {author} {\bibinfo {author} {\bibfnamefont {R.}~\bibnamefont
  {Clausius}},\ }\href
  {https://books.google.fr/books?id=8LIEAAAAYAAJ&printsec=frontcover&hl=fr&source=gbs_ge_summary_r&cad=0#v=onepage&q&f=false}
  {\emph {\bibinfo {title} {The mechanical theory of heat}}}\ (\bibinfo
  {publisher} {Macmillan \& Co, London, UK},\ \bibinfo {year}
  {1879})\BibitemShut {NoStop}%
\bibitem [{\citenamefont {Hecht}(2019)}]{Hecht_2019b}%
  \BibitemOpen
  \bibfield  {author} {\bibinfo {author} {\bibfnamefont {E.}~\bibnamefont
  {Hecht}},\ }\bibfield  {title} {\bibinfo {title} {Understanding energy as a
  subtle concept: A model for teaching and learning energy},\ }\href
  {https://doi.org/10.1119/1.5109863} {\bibfield  {journal} {\bibinfo
  {journal} {American Journal of Physics}\ }\textbf {\bibinfo {volume} {87}},\
  \bibinfo {pages} {495} (\bibinfo {year} {2019})}\BibitemShut {NoStop}%
\bibitem [{\citenamefont {Hecht}(2016)}]{Hecht_2016}%
  \BibitemOpen
  \bibfield  {author} {\bibinfo {author} {\bibfnamefont {E.}~\bibnamefont
  {Hecht}},\ }\bibfield  {title} {\bibinfo {title} {Relativity, potential
  energy, and mass},\ }\href {https://doi.org/10.1088/0143-0807/37/6/065804}
  {\bibfield  {journal} {\bibinfo  {journal} {European Journal of Physics}\
  }\textbf {\bibinfo {volume} {37}},\ \bibinfo {pages} {065804} (\bibinfo
  {year} {2016})}\BibitemShut {NoStop}%
\bibitem [{\citenamefont {de~Gennes}(1979)}]{deGennes_1979}%
  \BibitemOpen
  \bibfield  {author} {\bibinfo {author} {\bibfnamefont {P.-G.}\ \bibnamefont
  {de~Gennes}},\ }\href@noop {} {\emph {\bibinfo {title} {Scaling concepts in
  polymer physics}}}\ (\bibinfo  {publisher} {Cornell Univ. Press},\ \bibinfo
  {year} {1979})\BibitemShut {NoStop}%
\bibitem [{\citenamefont {Brillouin}(1965{\natexlab{a}})}]{Brillouin_1965a}%
  \BibitemOpen
  \bibfield  {author} {\bibinfo {author} {\bibfnamefont {L.}~\bibnamefont
  {Brillouin}},\ }\bibfield  {title} {\bibinfo {title} {The actual mass of
  potential energy, a correction to classical relativity},\ }\href
  {https://doi.org/10.1073/pnas.53.3.475} {\bibfield  {journal} {\bibinfo
  {journal} {Proceedings of the National Academy of Sciences}\ }\textbf
  {\bibinfo {volume} {53}},\ \bibinfo {pages} {475} (\bibinfo {year}
  {1965}{\natexlab{a}})}\BibitemShut {NoStop}%
\bibitem [{\citenamefont {Brillouin}(1965{\natexlab{b}})}]{Brillouin_1965b}%
  \BibitemOpen
  \bibfield  {author} {\bibinfo {author} {\bibfnamefont {L.}~\bibnamefont
  {Brillouin}},\ }\bibfield  {title} {\bibinfo {title} {The actual mass of
  potential energy ii},\ }\href {https://doi.org/10.1073/pnas.53.6.1280}
  {\bibfield  {journal} {\bibinfo  {journal} {Proceedings of the National
  Academy of Sciences}\ }\textbf {\bibinfo {volume} {53}},\ \bibinfo {pages}
  {1280} (\bibinfo {year} {1965}{\natexlab{b}})}\BibitemShut {NoStop}%
\bibitem [{\citenamefont {Szilard}(1964)}]{Szilard_1964}%
  \BibitemOpen
  \bibfield  {author} {\bibinfo {author} {\bibfnamefont {L.}~\bibnamefont
  {Szilard}},\ }\bibfield  {title} {\bibinfo {title} {On the decrease of
  entropy in a thermodynamic system by the intervention of intelligent
  beings},\ }\href {https://doi.org/10.1002/bs.3830090402} {\bibfield
  {journal} {\bibinfo  {journal} {Behavioral Science}\ }\textbf {\bibinfo
  {volume} {9}},\ \bibinfo {pages} {301} (\bibinfo {year} {1964})}\BibitemShut
  {NoStop}%
\bibitem [{\citenamefont {Feynman}\ \emph {et~al.}(1966)\citenamefont
  {Feynman}, \citenamefont {Leighton},\ and\ \citenamefont
  {Sands}}]{Feynmann_Ratchet}%
  \BibitemOpen
  \bibfield  {author} {\bibinfo {author} {\bibfnamefont {R.~P.}\ \bibnamefont
  {Feynman}}, \bibinfo {author} {\bibfnamefont {R.~B.}\ \bibnamefont
  {Leighton}},\ and\ \bibinfo {author} {\bibfnamefont {M.}~\bibnamefont
  {Sands}},\ }\href {https://www.feynmanlectures.caltech.edu/I_46.html} {\emph
  {\bibinfo {title} {The {F}eynman lectures on physics}}}\ (\bibinfo
  {publisher} {Addison-Wesley, Reading, MA},\ \bibinfo {year} {1966})\
  Chap.~\bibinfo {chapter} {46}, pp.\ \bibinfo {pages} {1--9}\BibitemShut
  {NoStop}%
\bibitem [{\citenamefont {Brillouin}(1950)}]{Brillouin1950}%
  \BibitemOpen
  \bibfield  {author} {\bibinfo {author} {\bibfnamefont {L.}~\bibnamefont
  {Brillouin}},\ }\bibfield  {title} {\bibinfo {title} {Can the rectifier
  become a thermodynamical demon?},\ }\href
  {https://doi.org/10.1103/physrev.78.627.2} {\bibfield  {journal} {\bibinfo
  {journal} {Physical Review}\ }\textbf {\bibinfo {volume} {78}},\ \bibinfo
  {pages} {627} (\bibinfo {year} {1950})}\BibitemShut {NoStop}%
\bibitem [{\citenamefont {Bormashenko}(2019)}]{Bormashenko_2019b}%
  \BibitemOpen
  \bibfield  {author} {\bibinfo {author} {\bibfnamefont {E.}~\bibnamefont
  {Bormashenko}},\ }\bibfield  {title} {\bibinfo {title} {The {L}andauer
  principle: Re-formulation of the second thermodynamics law or a step to great
  unification?},\ }\href {https://doi.org/10.3390/e21100918} {\bibfield
  {journal} {\bibinfo  {journal} {Entropy}\ }\textbf {\bibinfo {volume} {21}},\
  \bibinfo {pages} {918} (\bibinfo {year} {2019})}\BibitemShut {NoStop}%
\bibitem [{\citenamefont {Jaynes}(1957)}]{Jaynes_1957}%
  \BibitemOpen
  \bibfield  {author} {\bibinfo {author} {\bibfnamefont {E.~T.}\ \bibnamefont
  {Jaynes}},\ }\bibfield  {title} {\bibinfo {title} {Information theory and
  statistical mechanics},\ }\href {https://doi.org/10.1103/PhysRev.106.620}
  {\bibfield  {journal} {\bibinfo  {journal} {Phys. Rev.}\ }\textbf {\bibinfo
  {volume} {106}},\ \bibinfo {pages} {620} (\bibinfo {year}
  {1957})}\BibitemShut {NoStop}%
\bibitem [{\citenamefont {Jaynes}(1968)}]{Jaynes_1968}%
  \BibitemOpen
  \bibfield  {author} {\bibinfo {author} {\bibfnamefont {E.~T.}\ \bibnamefont
  {Jaynes}},\ }\bibfield  {title} {\bibinfo {title} {Prior probabilities},\
  }\href {https://doi.org/10.1109/TSSC.1968.300117} {\bibfield  {journal}
  {\bibinfo  {journal} {IEEE Transactions on Systems Science and Cybernetics}\
  }\textbf {\bibinfo {volume} {4}},\ \bibinfo {pages} {227} (\bibinfo {year}
  {1968})}\BibitemShut {NoStop}%
\bibitem [{\citenamefont {Shore}\ and\ \citenamefont
  {Johnson}(1980)}]{Shore_1980}%
  \BibitemOpen
  \bibfield  {author} {\bibinfo {author} {\bibfnamefont {J.}~\bibnamefont
  {Shore}}\ and\ \bibinfo {author} {\bibfnamefont {R.}~\bibnamefont
  {Johnson}},\ }\bibfield  {title} {\bibinfo {title} {Axiomatic derivation of
  the principle of maximum entropy and the principle of minimum
  cross-entropy},\ }\href {https://doi.org/10.1109/tit.1980.1056144} {\bibfield
   {journal} {\bibinfo  {journal} {{IEEE} Transactions on Information Theory}\
  }\textbf {\bibinfo {volume} {26}},\ \bibinfo {pages} {26} (\bibinfo {year}
  {1980})}\BibitemShut {NoStop}%
\bibitem [{\citenamefont {Lairez}(2023{\natexlab{b}})}]{Lairez_2023c}%
  \BibitemOpen
  \bibfield  {author} {\bibinfo {author} {\bibfnamefont {D.}~\bibnamefont
  {Lairez}},\ }\href {https://doi.org/10.48550/ARXIV.2310.17665} {\bibinfo
  {title} {Thermostatistics, information, subjectivity, why is this association
  so disturbing?}} (\bibinfo {year} {2023}{\natexlab{b}})\BibitemShut {NoStop}%
\end{thebibliography}%
	
\end{document}